\documentclass[aps,twocolumn,raggedbottom,prb,showpacs,nobalancelastpage,amssymb,groupedaddress]{revtex4}
\usepackage{graphicx}
\usepackage{amsmath}
\usepackage{amssymb}
\newcommand{\bra}[1]{\langle #1|}	%
\newcommand{\ket}[1]{|#1\rangle}

\usepackage{dsfont}

\begin{document}
\title{
Dynamics of a qubit coupled to a dissipative nonlinear quantum oscillator: an effective bath approach}
\author{Carmen\, Vierheilig$^1$, Dario Bercioux$^2$ and Milena\, Grifoni$^1$}
\affiliation{$^1$ Institut f\"{u}r Theoretische Physik, Universit\"at
Regensburg, 93035 Regensburg, Germany\\
$^2$ Freiburg Institute for Advanced Studies, Albert-Ludwigs-Universit\"at Freiburg,
79104 Freiburg im Breisgau, Germany}
\date{\today}

\begin{abstract}
We consider a qubit coupled to a nonlinear quantum oscillator, the latter coupled to an Ohmic bath, and investigate the qubit dynamics. This composed system can be mapped onto that of a qubit coupled to an effective bath. An approximate mapping procedure to determine the spectral density of the effective bath is given. Specifically, within a linear response approximation the effective spectral density is given by the know\-ledge of the linear susceptibility of the nonlinear quantum oscillator. To determine the actual form of the susceptibility, we consider its periodically driven counterpart, the problem of the quantum Duffing oscillator within linear response theory in the driving amplitude. Knowing the effective spectral density, the qubit dynamics is investigated. In particular, an analytic formula for the qubit's population difference is derived. Within the regime of validity of our theory, a very good agreement is found with predictions obtained from a Bloch-Redfield master equation approach applied to the composite qubit-nonlinear oscillator system.
\end{abstract}
 \maketitle

\section{Introduction}
The understanding of relaxation and dephasing properties of qubits due to the surrounding environment is essential for quantum computation \cite{Nielsen}. A famous model to study the environmental influences on the coherent dynamics of a qubit is the spin-boson model \cite{Weiss,LeggettRevModPhysErratum,Grifoni304}, consisting of a two-level system (TLS) bilinearily coupled to a bath of harmonic oscillators. Although the bath degrees of freedom can be traced out exactly, analytical solutions are only possible within perturbative schemes.
First those perturbative in the coupling of the TLS to the bath are typically obtained within a Born-Markov treatment of the Liouville equation for the TLS density matrix \cite{Redfield,Blum} or within the path integral formalism \cite{Weiss}. The equivalence of both methods has been demonstrated restricting to low temperatures and low damping strengths in Ref. [\onlinecite{Hartmann}]. The second alternative approach is to perform perturbation theory in the tunneling amplitude of the two-level system. Within the so-termed non-interacting blip approximation (NIBA) \cite{LeggettRevModPhysErratum,Grifoni304,Weiss} it yields equations of motion for the TLS reduced density matrix enabling to capture the case of strong TLS-bath coupling. Reality is however often more complex, as the qubit might be coupled to other quantum systems besides to a thermal bath. For example, to read-out its state, a qubit is usually coupled to a read-out device.\\
In the following we mostly have in mind the flux qubit \cite{Mooij1999} read-out by a DC-SQUID. The latter mediates the dissipation originating from the surrounding electromagnetic bath and can be modeled both as a linear or nonlinear oscillator \cite{Mooij1999,Tian,vanderWal,Chiorescu2003,Chiorescu2004,Johansson2006,Lupascuquantumstatedetection,Lee,Picot,LupascuHigh,Inomata,Wirth}. Recently, the nonlinearity of qubit read-out devices, for example of a DC-SQUID \cite{Lee,Picot,LupascuHigh,Inomata,Wirth} or a Josephson bifurcation amplifier \cite{Siddiqi2,Siddiqi,MalletSingle}, has been used to improve the measurement scheme in terms of a faster read-out and higher fidelity. Specifically, the device was operated in a regime where the dynamics exhibited bifurcation features typical of a classical nonlinear oscillator. As demonstrated e.g. in Ref. [\onlinecite{Chiorescu2004}] the quantum limit is within the experimental reach as well.\\
From the theoretical side there are two different viewpoints to investigate the dynamics of a qubit coupled to an oscillator, with the latter in turn coupled to a thermal bath. The first way is to consider the TLS and the oscillator as a single quantum system coupled to the bath, while the second is an effective bath description where the effective environment seen by the qubit includes the oscillator and the original thermal bath. The mapping to an effective bath has been discussed for the case in which the TLS is coupled to a \textit{harmonic} oscillator in Ref. [\onlinecite{Garg}]. Specifically, the spectral density of the effective bath acquires a broadened peak centered around the frequency of the oscillator. This case has been investigated in Refs. [\onlinecite{Wilhelm1,Goorden2005,Goorden2004,Goorden2003,Nesi,Kleff2,Brito}] by applying standard numerical and analytical methods established for the spin-boson model. All those works showed that the peaked structure of the effective bath is essential when the eigenfrequency of the TLS becomes comparable to the oscillator frequency.\\
So far the first approach was used in Ref. [\onlinecite{CarmenPRA}] to describe a qubit-nonlinear oscillator (NLO) system in the deep quantum regime. Here the effects of the (harmonic) thermal reservoir can be treated using standard Born-Markov perturbation theory. The price to be paid, however, is that the Hilbert space of the qubit-nonlinear oscillator system is infinite, which requires for practical calculations its truncation invoking e.g. low temperatures \cite{CarmenPRA}.\\
In contrast to the above work we investigate here the case of a qubit-NLO system, with the latter being coupled to an Ohmic bath, within an effective bath description. Due to the nonlinearity of the oscillator, the mapping to a \textit{linear} effective bath is not exact. In this case a temperature and nonlinearity dependent effective spectral density well captures the NLO influence on the qubit dynamics.\\
The paper is organized as follows: In section \ref{model} we introduce the model with the relevant quantities. In section \ref{mapping} the mapping procedure is investigated and the effective spectral density for the corresponding linear case is given. Afterwards the mapping procedure is applied to the case of a qubit coupled to a nonlinear quantum oscillator. As a consequence of the mapping the determination of the effective spectral density is directly related to the knowledge of the susceptibility of the oscillator. We show how the susceptibility can be obtained from the steady-state response of a quantum Duffing oscillator in section \ref{suslin}. In section \ref{susnonlin} the steady-state response of the dissipative quantum Duffing oscillator is reviewed and its susceptibility is put forward. The related effective spectral density is derived in section \ref{Jeff}. In section \ref{qudyn} the qubit dynamics is investigated by applying the non-interacting blip approximation (NIBA) to the kernels of the generalized master equation which governs the dynamics of the population difference of the qubit. A comparison with the results of Ref. [\onlinecite{CarmenPRA}], obtained within the first approach, is shown. Further, analogies and differences with respect to the linear case are discussed. In section \ref{Conclusions} conclusions are drawn.

\section{Hamiltonian}\label{model}
\begin{figure}[h!]
\begin{center}
\includegraphics[width=0.45\textwidth]{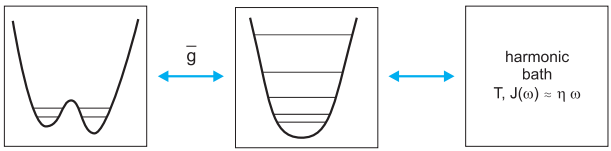}
\end{center}
\caption{Schematic representation of the composed system built of a qubit, an intermediate nonlinear oscillator and an Ohmic bath.
\label{linearbath}}
\end{figure}

\begin{figure}[h!]
\begin{center}
\includegraphics[width=0.45\textwidth]{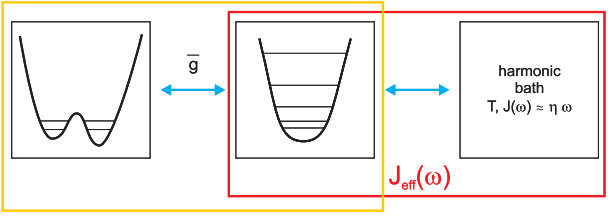}
\end{center}
\caption{Schematic representation of the complementary approaches available to evaluate the qubit dynamics: In the first approach one determines the eigenvalues and eigenfunctions of the composite qubit plus oscillator system (yellow (light grey) box) and accounts afterwards for the harmonic bath characterized by the Ohmic spectral density $J(\omega)$. In the effective bath description one considers an environment built of the harmonic bath and the nonlinear oscillator (\rm red (dark grey) box). In the harmonic approximation the effective bath is fully characterized by its effective spectral density $J_{\rm eff}(\omega)$.
\label{approachschaubild}}
\end{figure}

We consider a composed system built of a qubit, -the system of interest-, coupled to a nonlinear quantum oscillator (NLO), see Fig. \ref{linearbath}.
To read-out the qubit state we couple the qubit linearly to the oscillator with the coupling constant $\overline{g}$, such that via the intermediate NLO dissipation also enters the qubit dynamics.
The Hamiltonian of the composed system reads:
\begin{equation}\label{eq10}
\hat{H}_{\rm tot}=\hat{H}_{\rm S}+\hat{H}_{\rm NLO}+\hat{H}_{\rm S+NLO}+\hat{H}_{\rm NLO+B}+\hat{H}_{\rm B},
\end{equation}
where
\begin{eqnarray}\label{gl3}
\hat{H}_{\rm S}&=&\frac{\hat{p}^2}{2\mu}+U(\hat{q}),\\
\hat{H}_{\rm NLO}&=& \frac{1}{2M}\hat{P}_y^2+\frac{1}{2} M\Omega^2\hat{y}^2+\frac{\overline{\alpha}}{4}\hat{y}^4,\nonumber\\
\hat{H}_{\rm S+NLO}&=&\overline{g}\hat{y}\hat{q},\nonumber\\
\hat{H}_{\rm NLO+B}&=& \sum_j\left[-c_j \hat{x}_j \hat{y}+\frac{c_j^2}{2m_j\omega_j^2}\hat{y}^2\right],\nonumber\\
\hat{H}_{\rm B}&=&\sum_j\left[
\frac{\hat{p}_j^2}{2m_j}+\frac{1}{2}m_j \omega_j^2\hat{x}_j^2\right]\nonumber.
\end{eqnarray}
Here $\hat{H}_{\rm S}$ represents the qubit Hamiltonian, where $\mu$ is the particle's mass and $U(q)$ a one-dimensional double well  potential with minima at $q=\pm q_0/2$. $\hat{H}_{\rm NLO}$ is the NLO Hamiltonian, where the parameter $\overline{\alpha}>0$ accounts for the nonlinearity. When the oscillator represents a SQUID used to read-out the qubit, the oscillator frequency $\Omega$ corresponds to the SQUID's plasma frequency. The dissipation on the NLO is modeled in the following by coupling it to an Ohmic bath, characterized by the spectral density \cite{Weiss}:
\begin{equation}\label{ohmic}
J(\omega)=\frac{\pi}{2}\sum_{j=1}^\mathcal{N}\frac{c_j^2}{m_j\omega_j}\delta(\omega-\omega_j)=\eta\omega=M\gamma\omega.
\end{equation}
In the classical limit it corresponds to a white noise source, where $\eta$ is a friction coefficient with dimensions mass times frequency.\\
In the following focus will be on the qubit dynamics in the presence of the dissipative nonlinear oscillator. Namely we will study the time evolution of the qubit's position as described by:
\begin{eqnarray}
q(t)&:=&{\rm Tr}\{\hat{\rho}_{\rm tot}(t)\hat{q}\}={\rm Tr}_{\rm S}\{\hat{\rho}_{\rm red}(t)\hat{q}\},
\end{eqnarray}
where $\hat{\rho}_{\rm tot}$ and $\hat{\rho}_{\rm red}$ are the total and reduced density operators, respectively. The latter is defined as:
\begin{eqnarray}
\hat{\rho}_{\rm red}:={\rm Tr}_{\rm B} {\rm Tr}_{\rm NLO}\{\hat{\rho}_{\rm tot}\},
\end{eqnarray}
where the trace over the degrees of freedom of the bath and of the oscillator is taken. In Fig. \ref{approachschaubild} two different approaches to determine the qubit dynamics are depicted. In the first approach, which is elaborated in Ref. [\onlinecite{CarmenPRA}], one first determines the eigenstates and eigenvalues of the composed qubit-oscillator system and then  includes environmental effects via standard Born-Markov perturbation theory. The second approach exploits an effective description for the environment surrounding the qubit based on a mapping procedure. This will be investigated in the next section.
\section{Mapping}\label{mapping}
The main aim is to evaluate the qubit's evolution described by $q(t)$. This can be achieved within an effective description using a mapping procedure. Thereby the oscillator and the Ohmic bath are put together, as depicted in Figure \ref{approachschaubild}, to form an effective bath. The effective Hamiltonian
\begin{eqnarray}\label{gl1}
\hat{H}_{\rm eff}&=&\hat{H}_{\rm S}+\hat{H}_{\rm B eff}
\end{eqnarray}
is chosen such that, after tracing out the bath degrees of freedom, the same dynamical equations for $q(t)$ are obtained as from the original Hamiltonian $\hat{H}_{\rm tot}$. Due to the nonlinear character of the oscillator an exact mapping implies that $\hat{H}_{\rm B eff}$  represents a nonlinear environment. We shall show in the following subsection using linear response theory that a linear approximation for $\hat{H}_{\rm B eff}$ is justified for weak coupling $\overline{g}$. Then Eq. (\ref{gl1}) describes an effective spin-boson problem where
\begin{eqnarray}\label{effbath}
\hat{H}_{\rm B eff}&=&\frac{1}{2}\sum_{j=1}^\mathcal{N}\left[\frac{\hat{P}_j^2}{m_j}+m_j\omega_j^2\left(\hat{X}_j-
\frac{d_j}{m_j \omega_j^2} \hat{q}\right)^2\right],
\end{eqnarray}
and the associated spectral density is:
\begin{equation}
J_{\rm{eff}}(\omega)=\frac{\pi}{2}\sum_{j=1}^\mathcal{N}\frac{d_j^2}{m_j\omega_j}\delta(\omega-\omega_j).
\end{equation}
The Hamiltonian (\ref{gl1}) with (\ref{effbath}) leads to coupled equations of motion \cite{Weiss,LeggettRevModPhysErratum}:
\begin{eqnarray}
\mu\ddot{\hat{q}}(t)+U'(\hat{q})+\sum_{j=1}^\mathcal{N}\left(\frac{d_j^2}{m_j\omega_j^2}\hat{q}\right)&=&\sum_{j=1}^\mathcal{N} d_j \hat{X}_j,\nonumber\\
m_j \ddot{\hat{X}}_j+m_j\omega_j^2\hat{X}_j&=&d_j \hat{q}\nonumber,
\end{eqnarray}
where $U'(\hat{q})=\frac{d}{dq}U(\hat{q})$. By formally integrating the second equation of motion and inserting the solution into the first equation the well-known Langevin equation for the operator $\hat{q}$ is obtained. This, in turn, allows to obtain the Langevin equation for $q_{\rm eff}(t):={\rm Tr}\{\hat{\rho}_{\rm eff}\hat{q}(t)\}$\cite{Weiss}:
\begin{eqnarray}\label{eom1}
\mu \ddot{q}_{\rm eff}+\mu\int_{0}^t dt'\gamma_{\rm eff}(t-t')\dot{q}_{\rm eff}+\langle U'(\hat{q})\rangle_{\rm eff}&=&0,
\end{eqnarray}
with the effective damping kernel $\gamma_{\rm eff}(t-t')$.\\
Notice that $\langle\dots\rangle_{\rm eff}$ indicates the expectation value taken with respect to $\hat{\rho}_{\rm{eff}}$, which is the density operator associated to $\hat{H}_{\rm{eff}}$ \cite{Weiss}. In Laplace space, defined by
\begin{eqnarray}
y(t)&=&\frac{1}{2\pi i}\int_{\mathcal{C}} d\lambda y(\lambda)\exp(\lambda t),\\
y(\lambda)&=&\int_{0}^\infty dt y(t)\exp(-\lambda t)\nonumber,
\end{eqnarray}
we obtain from Eq. (\ref{eom1}) the equation of motion:
\begin{equation}\label{map2}
\mu \lambda^2 q_{\rm eff}(\lambda)+\mu \lambda\gamma_{\rm eff}(\lambda) q_{\rm eff}(\lambda)+\langle U'(\lambda)\rangle_{\rm eff}=0.
\end{equation}
The real part $\gamma'_{\rm{eff}}(\omega)=\textrm{Re}[\hat{\gamma}(\lambda=-i\omega)]$ of the effective damping kernel $\gamma_{\rm eff}(t)$ is related to the spectral density via \cite{Weiss}:
\begin{eqnarray}\label{gammaJ}
\gamma'_{\rm{eff}}(\omega)=\frac{J_{\rm eff}(\omega)}{\mu\omega}.
\end{eqnarray}
The mapping for the case of zero nonlinearity $\overline{\alpha}$ and Ohmic damping has been discussed in Ref. [\onlinecite{Garg}]. There the influence of both the intermediate harmonic oscillator and the bath is embedded into an effective peaked spectral density given by:
\begin{eqnarray}\label{linearspecdens}
 J_{\rm{eff}}^{\rm HO} (\omega)&=&\frac{\overline{g}^2\gamma\omega}{M(\Omega^2 - \omega^2)^2 + M\gamma^2 \omega^2 },
\end{eqnarray}
showing Ohmic low frequency behaviour $J_{\textrm{eff}}^{HO} (\omega)\longrightarrow_{\ _{\hspace{-0.7cm}\omega\rightarrow 0}} \overline{g}^2\gamma\omega/(M\Omega^4)$.

\subsection{Equation of motion for the nonlinear Hamiltonian}
As discussed above, the mapping requires the knowledge of the \textit{reduced} dynamics of the system described by the variable $q(t)$.
Therefore we start from the coupled equations of motion derived from the Hamiltonian $\hat{H}_{\rm tot}$ given in Eq. (\ref{eq10}):
\begin{subequations}
\begin{eqnarray}
\mu \ddot{\hat{q}}+U'(\hat{q})=-\overline{g}\hat{y}\label{eq7a},\\
M\ddot{\hat{y}}+\eta\dot{\hat{y}}+M\Omega^2 \hat{y}+\overline{\alpha} \hat{y}^3&=&-\overline{g}\hat{q}+\hat{\xi}(t).\label{eq7b}
\end{eqnarray}
\end{subequations}
According to Eq. (\ref{ohmic}), $\eta=M\gamma$ is the damping coefficient and
\begin{equation}
\hat{\xi}(t)=\sum_{j=1}^\mathcal{N} c_j\left[x_j^{(0)}\cos(\omega_j t)+\frac{p_j^{(0)}}{m_j\omega_j}\sin(\omega_j t)\right]-M\gamma\delta(t)\hat{y}(0)
\end{equation}
a fluctuating force originating from coupling to the bath. In order to eliminate $\hat{y}$ from the first equation of motion, we have to calculate $\hat{y}[\hat{q}(t)]$ from the second equation.\\
In the following we look at equations of motion for the expectation values resulting from Eqs. (\ref{eq7a}) and (\ref{eq7b}), i.e., we look at the evolution of $q(t):={\rm Tr}\{\hat{\rho}_{\rm tot}\hat{q}(t)\}$ and $y(t):={\rm Tr}\{\hat{\rho}_{\rm tot}\hat{y}(t)\}$. Since we want to calculate $y(t)$ we turn back to Eq. (\ref{eq10}) and treat the coupling term $\hat{H}_{\rm S+NLO}$ as a perturbation, $\overline{g}\ll M\Omega^2$. Then the use of linear response theory in this perturbation is justified and we find:\\
\begin{eqnarray}\label{eq1}
\lefteqn{y(t)=\langle\hat{y}(t)\rangle_0}\\ &&-\frac{i}{\hbar}\int_{-\infty}^{\infty}dt'\theta(t-t')\langle[\hat{y}(t),\hat{y}(t')]\rangle_0 \overline{g} \langle \hat{q}(t')\rangle_0\theta(t')\nonumber\\
&&+\mathcal{O}(\overline{\alpha}\,\overline{ g}^2),\nonumber
\end{eqnarray}
where $\langle\dots\rangle_0$ denotes the expectation value in the absence of the coupling $\overline{g}$, which we assume it has been switched on at time $t_0=0$.\\
Notice that for a \textit{linear} system, as for example the damped harmonic oscillator, the linear response \textit{becomes exact}, such that the neglected corrections are at least of order $\mathcal{O}(\overline{\alpha}\,\overline{ g}^2)$. Moreover, the time evolution of the expectation values is the same as in \textit{the classical case}; this fact corresponds to the Ehrenfest theorem \cite{Weiss}. For \textit{nonlinear} systems the expression in Eq. (\ref{eq1}) is an approximation, because all orders in the perturbation are nonvanishing\begin{footnote}{
An extension of the concept of linear response in case of nonlinear systems is the so called Volterra expansion, which provides a systematic perturbation series in the forcing \cite{Volterra}.}
\end{footnote}.\\
In Laplace space,
Eq. (\ref{eq1}) yields:\\
\begin{eqnarray}\label{f1}
\delta y(\lambda)=\chi(\lambda)\overline{g}\langle \hat{q}(\lambda)\rangle_0+\mathcal{O}(\overline{\alpha}\,\overline{ g}^2),
\end{eqnarray}
where $\delta y(\lambda)=y(\lambda)-\langle \hat{y}(\lambda)\rangle_0$ and
where $\chi(\lambda)$ is the Laplace transform of the response function or susceptibility:\\
\begin{eqnarray}
\chi(t-t')=-\frac{i}{\hbar}\theta(t-t')\langle[\hat{y}(t),\hat{y}(t')]\rangle_0.
\end{eqnarray}
Since $q(\lambda)-\langle \hat{q}(\lambda)\rangle_0=\mathcal{O}(\overline{g}^2)$,
from Eqs. (\ref{eq7a}) and (\ref{f1}) it follows:\\
\begin{eqnarray}
&&\mu \lambda^2q(\lambda)+\overline{g}^2\chi(\lambda)q(\lambda)+\mathcal{O}(\overline{\alpha}\,\overline{g}^3,\overline{g}^4)\label{eqneu}\nonumber\\
&&=-\langle U'(\lambda)\rangle-\overline{g}\langle \hat{y}(\lambda)\rangle_0.
\end{eqnarray}
That is, we have a normalization of the mass, and a damping-like term due to the coupled equations of motion. The effect of the nonlinearity is embedded in the response function $\chi$.\\
We assume in the following that in the absence of the coupling to the qubit the NLO and bath are in thermal equilibrium, which yields $\langle\hat{y}(t)\rangle_0=0$ for all times, and thus also: $\langle\hat{y}(\lambda)\rangle_0=0$.

\subsection{Mapping of the equations of motion and generic form for the effective spectral density}
By comparison of Eqs. (\ref{map2}) and (\ref{eqneu}) we can conclude that they yield the same dynamics if:
\begin{eqnarray}\label{rel1}
\langle U'(\lambda)\rangle_{\rm eff}=\langle U'(\lambda)\rangle,
\end{eqnarray}
and the effective bath is chosen such that:
\begin{eqnarray}
\overline{g}^2\frac{\chi(\lambda)}{\mu \lambda}&=&\gamma_{\rm eff}(\lambda).
\end{eqnarray}
By comparing the last equations with the relation (\ref{gammaJ}) and replacing $\lambda=-i\omega$
it follows:
\begin{eqnarray}\label{gl20}
J_{\rm eff}(\omega)= -\overline{g}^2\chi''(\omega),
\end{eqnarray}
where $\chi''(\omega)$ is the imaginary part of the susceptibility in Fourier space. We have now reduced the problem of finding the effective spectral density to that of determining the corresponding susceptibility. Notice that for a linear system the classical and quantum susceptibility coincide and are independent of the driving amplitude! In this case it is possible to calculate $\chi(\omega)$  directly from the classical equations of motion. For a generic nonlinear system, however, the classical and quantum susceptibilities differ.

\subsection{Linear susceptibility of a Duffing oscillator}\label{suslin}
In order to evaluate the linear susceptibility, we solve the auxiliary problem of calculating the susceptibility of a quantum Duffing oscillator (DO), i.e., of the nonlinear quantum oscillator in Eq. (\ref{gl3}) additionally driven by a periodic force with driving amplitude $F$ and driving frequency $\omega_{ex}$.
The corresponding equation of motion is:
\begin{equation}
M\ddot{\hat{y}}+\eta\dot{\hat{y}}+M\Omega^2 \hat{y}+\overline{\alpha} \hat{y}^3=-F\theta(t-t_0)\cos(\omega_{ex} t)+\hat{\xi}(t).
\end{equation}
Application of linear response theory in the driving yields the equation for the expectation value of the position of the oscillator:
\begin{eqnarray}
y(t)&=& \langle \hat{y}(t)\rangle_0-\frac{i}{\hbar}\int_{t_0}^{\infty}dt'\theta(t-t')\langle[\hat{y}(t),\hat{y}(t')]\rangle_0\nonumber\\
&&\times F\cos(\omega_{ex} t') +\mathcal{O}(F^2).
\end{eqnarray}
Using the symmetry properties of the susceptibility $\chi(\omega)$ we obtain in the steady-state limit:
\begin{eqnarray}\label{chisteadystate}
y_{st}(t)=\lim_{t_0\rightarrow -\infty}y(t)&=& \langle \hat{y}(t)\rangle_0+F\cos(\omega_{ex} t)\chi'(\omega_{ex})\nonumber\\
&&+ F\sin(\omega_{ex} t)\chi''(\omega_{ex})\nonumber+\mathcal{O}(F^3)\\
&\equiv&A\cos(\omega_{ex}t+\phi)+\mathcal{O}(F^3).
\end{eqnarray}
Here the presence of the Ohmic bath implies $\lim_{t_0\rightarrow -\infty}\langle \hat{y}(t)\rangle_0=0$. Notice that due to symmetry inversion of the NLO, corrections of $\mathcal{O}(F^2)$ vanish in Eq. (\ref{chisteadystate}).
In Eq. (\ref{chisteadystate}) $A$ and $\phi$ are the amplitude and phase of the steady-state response. It follows $\chi(\omega)=\frac{A}{F}\exp(-i\phi)$, such that $\chi''(\omega)=-\frac{A}{F}\sin\phi$.

\section{Steady-state dynamics of a Duffing oscillator}\label{susnonlin}
So far we have reduced the problem of finding the effective spectral density to the one of determining the steady-state response of the Duffing oscillator in terms of the amplitude $A$ and the phase $\phi$.
These quantities were recently derived in Refs. [\onlinecite{Peano2,CarmenChemPhys}], using the framework of a Bloch-Redfield-Floquet description of the dynamics of the DO. The results in Ref. [\onlinecite{CarmenChemPhys}] are applicable in a wide range of driving frequencies around the one-photon resonance regime $\omega_{ex}=\Omega+3\overline{\alpha} y_0^4/(4\hbar)\equiv\Omega_1$ for strong enough nonlinearities: $y_0 F/(2\sqrt{2})\ll 3\overline{\alpha} y_0^4/4\ll\hbar\Omega$, where we introduced the oscillator length $y_0=\sqrt{\hbar/(M\Omega)}$.\\
As illustrated in Refs. [\onlinecite{Peano2,CarmenChemPhys}] the amplitude and phase are fully determined by the knowledge of the matrix elements of the stationary density matrix of the Duffing oscillator in the Floquet basis, see e.g. Eqs. (67)-(70) in Ref. [\onlinecite{CarmenChemPhys}]. There the master equation yielding the elements of the stationary density matrix is analytically solved in the low temperature regime $k_B T<<\hbar\Omega$ imposing a partial secular approximation, yielding Eq. (70) of Ref. [\onlinecite{CarmenChemPhys}], and restricting to spontaneous emission processes only. Here we follow the same line of reasoning as in Ref. [\onlinecite{CarmenChemPhys}] to evaluate the amplitude and phase: we impose the same partial secular approximation and consider low temperatures $k_B T <\hbar\Omega$. However, we include now both emission and absorption processes, i.e., we use the full dissipative transition rates as in Eq. (64) of Ref. [\onlinecite{CarmenChemPhys}].
The imaginary part of the linear susceptibility $\chi$ follows from the so obtained \textit{nonlinear} susceptibility $\chi_{NL}$ in the limit of vanishing driving amplitudes:
\begin{small}
\begin{eqnarray}\label{chilarger}
\lefteqn{\chi''(\omega_{ex})=\lim_{F\rightarrow 0}\chi''_{NL}(\omega_{ex})}\\
&=&-\frac{y_0^4J(\omega_{ex})n_1(0)^4\frac{2\Omega_1}{|\omega_{ex}|+\Omega_1}}{y_0^4J(\Omega_1)^2n_1(0)^4(2 n_{th}(\Omega_1)+1)^2+4\hbar^2(|\omega_{ex}|-\Omega_1)^2},\nonumber
\end{eqnarray}
\end{small}
where
\begin{eqnarray}
n_1(0)&=&\left[1-\frac{3}{8\hbar\Omega}\overline{\alpha} y_0^4\right].
\end{eqnarray}
For consistency also $n_1^4(0)$ has to be treated up to first order in $\overline{\alpha}$ only.\\
Moreover, we used the spectral density $J(\omega)=M\gamma\omega$ and the Bose function $n_{th}(\epsilon)=\left[\coth\left(\hbar\epsilon/(2k_B T)\right)-1\right]/2$, which determines the weight of the emission and absorption processes.\\

\section{Effective spectral density for a nonlinear system} \label{Jeff}
\begin{figure}[h!]
\begin{center}
\includegraphics[width=0.45\textwidth]{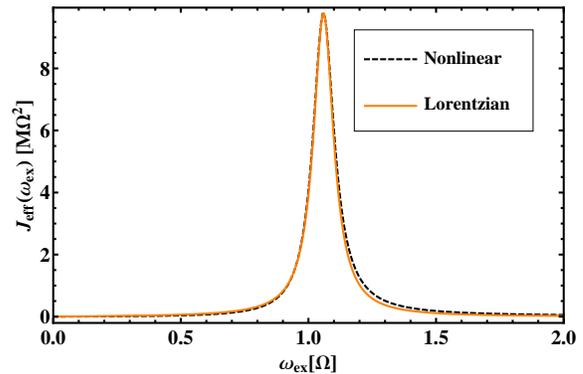}
\end{center}
\caption{Comparison of the effective spectral density $J_{\rm eff}(\omega)$ and a Lorentz curve for moderate damping. The parameters are: $\Omega=1.0$, $y_0^4\overline{\alpha}/(\hbar\Omega)=0.08$,  $\gamma=0.097\Omega$, and $\beta=10(\hbar\Omega)^{-1}$. \label{CompLorentz}}
\end{figure}

The effective spectral density follows from Eqs. (\ref{gl20}) and (\ref{chilarger}). It reads:
\begin{small}
\begin{eqnarray}\label{Jsimpl}
\lefteqn{
J_{\rm eff}(\omega_{ex})}\\
&&=\overline{g}^2\frac{\gamma\omega_{ex}n_1(0)^4\frac{2\Omega_1}{|\omega_{ex}|+\Omega_1}}{M\gamma^2\Omega_1^2(2 n_{th}(\Omega_1)+1)^2n_1(0)^4+4M\Omega^2(|\omega_{ex}|-\Omega_1)^2}.\nonumber
\end{eqnarray}
\end{small}
As in case of the effective spectral density $J_{\rm eff}^{\rm HO}$, Eq. (\ref{linearspecdens}), we observe Ohmic behaviour at low frequency. In contrast to the linear case, the effective spectral density is peaked at the shifted frequency $\Omega_1$. Its shape approaches the Lorentzian one of the linear effective spectral density, but with peak at the shifted frequency, as shown in Fig. \ref{CompLorentz}.\\
While in Refs. [\onlinecite{CarmenChemPhys,Peano2}] the amplitude of the oscillator showed an antiresonant to resonant transition depending on the ratio of driving amplitude $F$ and damping $\gamma$, the effective spectral density, obtained in the limit $F\to 0$,  displays only resonant behaviour.\\

\section{Qubit dynamics}\label{qudyn}
In the following we derive the dynamics of a qubit coupled to this effective nonlinear bath. Therefore we identify the system Hamiltonian $\hat{H}_{\rm S}$ introduced in Eq. (\ref{eq10}) with the one of a qubit, denoted in the following as $\hat{H}_{\rm TLS}$. This is verified at low energies if the barrier height of the double well potential $U(\hat{q})$ is larger than the energy separation of the ground and first excited levels in each well. In this case the relevant Hilbert space can be restricted to the two-dimensional space spanned by the ground state vectors $\ket{L}$ and $\ket{R}$ in the left and right potential well, respectively \cite{Weiss}. We start defining the actual form of the qubit Hamiltonian and its interaction with the nonlinear oscillator and afterwards introduce its dynamical quantity of interest, the population difference $P(t)$.
\subsection{Qubit}
The Hamiltonian of the TLS (qubit), given in the localized basis $\left\{|L\rangle,|R\rangle\right\}$, is:
\begin{eqnarray}
\hat{H}_{\rm TLS}&=&-\frac{\hbar}{2}\left(\varepsilon\sigma_z+\Delta\sigma_x\right),
\end{eqnarray}
where $\sigma_{i}$, $i=x,z$, are the corresponding Pauli matrices, the energy bias $\varepsilon$ accounts for an asymmetry between the two wells and $\Delta$ is the tunneling amplitude. The bias $\varepsilon$ can be tuned for a superconducting flux qubit by application of an external flux $\Phi_{\rm ext}$ and vanishes at the so-called degeneracy point \cite{WallraffSideband}. For $\varepsilon\gg\Delta$ the states $|L\rangle$ and $|R\rangle$ are eigenstates of $\hat{H}_{\rm TLS}$, corresponding to clockwise and counterclockwise currents, respectively.\\
The interaction in Eq. (\ref{gl3}) is conveniently rewritten as:
\begin{eqnarray}\label{parameter1}
\hat{H}_{\rm TLS-NLO}&=&\overline{g}\hat{q}\hat{y}\\
&=&\frac{\overline{g}}{2\sqrt{2}}q_0\sigma_z y_0(a+a^\dagger)\nonumber\\
&:=&\hbar g\sigma_z (a+a^\dagger).\nonumber
\end{eqnarray}
Likewise we express the nonlinear oscillator Hamiltonian as:
\begin{eqnarray}\label{parameter2}
\hat{H}_{\rm NLO}&=& \hbar\Omega\left(\hat{j}+\frac{1}{2}\right)+\frac{\overline{\alpha}}{4}\hat{y}^4\\
&=& \hbar\Omega\left(\hat{j}+\frac{1}{2}\right)+\frac{\overline{\alpha}y_0^4}{16}(a+a^\dagger)^4\nonumber\\
&:=&\hbar\Omega\left(\hat{j}+\frac{1}{2}\right)+\frac{\alpha}{4}(a+a^\dagger)^4.\nonumber
\end{eqnarray}

\subsection{Population difference}

The dynamics of a qubit is usually characterized in terms of the population difference $P(t)$ between the $\ket{R}$ and $\ket{L}$ states of the qubit:
\begin{eqnarray}\label{dynamics}
P(t)&:=&\langle\sigma_z\rangle\\
&=&{ \rm} {\rm Tr}_{{\rm TLS}} \{ \sigma_{ {\rm z}} \hat{\rho}_{{\rm \rm red}} (t) \}\\& =& \bra{{\rm R}}  \hat{\rho}_{{\rm  red}}(t) \ket{{\rm R}} - \bra{{\rm L}}  \hat{\rho}_{{\rm red}}(t) \ket{{\rm L}},\nonumber
\end{eqnarray}
where $\hat{\rho}_{{\rm red}}(t)$ is the reduced density matrix of the TLS,
\begin{equation}
  \hat{\rho}_{{\rm red}}(t) ={\rm Tr}_{B} \{\hat{ \rho}_{\rm eff}(t) \}.
\end{equation}
It is found after tracing out the degrees of freedom of the effective bath from the total density matrix
$\hat{\rho}_{\rm eff} (t) = \exp^{- \frac{{\rm i}}{\hbar} \hat{H}_{\rm eff} t} \hat{\rho}_{\rm eff} (0) \exp^{\frac{{\rm i}}{\hbar}  \hat{H}_{\rm  eff} t}$. It follows that in the two level approximation $q_{\rm eff}(t)=\frac{q_0}{2}P(t)$, where $q_{\rm eff}(t)$ is the position operator expectation value introduced in Sec. \ref{mapping}.\\
As we mapped the nonlinear system onto an effective spin-boson model, the evaluation of the population difference $P(t)$ of the TLS is possible using standard approximations developed for the spin-boson model \cite{Goorden2004,Goorden2005,Nesi}.
Assuming a factorized initial condition $\hat{\rho}_{\rm eff}(0)=\hat{\rho}_{\rm TLS}(0)\exp(-\beta\hat{H}_{\rm Beff}/Z)$,
the population difference $P(t)$ fulfills the generalized master equation (GME) \cite{Weiss,GrifoniExact}
\begin{equation}\label{GME}
\dot{P}(t)=-\int_0^t dt'[K^s(t-t')P(t')+K^a(t')],\quad t>0
\end{equation}
where $K^s(t-t')$ and $K^a(t-t')$ are symmetric and antisymmetric with respect to the bias, respectively. They are represented as a series in the tunneling amplitude. As an exact solution neither analytically nor numerically is available, due to the complicated form of the exact kernel, we impose in the following the so-called Non-Interacting Blip Approximation (NIBA) \cite{Weiss,LeggettRevModPhysErratum}. Applying NIBA corresponds to truncating the exact kernels to first order in $\Delta^2$ and is therefore perturbative in the tunneling amplitude of the qubit. It is justified in various regimes: it is exact at zero damping, otherwise it is only an approximation which works at best for zero bias and/or large damping and/or high temperature \cite{Weiss}. One finds within the NIBA
\begin{eqnarray}\label{kernels}
K^s(t)&=& \Delta^2\exp(-S(t))\cos(R(t)),\\
K^a(t)&=& \Delta^2\exp(-S(t))\sin(R(t)),\nonumber
\end{eqnarray}
where $S(\tau)$ and $R(\tau)$ are the real and imaginary part of the bath correlation function:
\begin{eqnarray}
Q(\tau)&=&S(\tau)+iR(\tau)=\int_0^\infty d\omega \frac{G_{\rm eff}(\omega)}{\omega^2}\times\\
&&\left[\coth\left(\frac{\beta\hbar\omega}{2}(1-\cos(\omega t))\right)+i\sin(\omega t)\right]\nonumber,
\end{eqnarray}
where $G_{\rm eff}(\omega)=q_0^2J_{\rm eff}(\omega)/(\pi\hbar)$. In particular, upon introducing the dimensionless constant \mbox{$\varsigma=g^2\gamma n_1(0)^4/(\pi\Omega^3)$}, we obtain:
\begin{eqnarray}\label{GeffNL}
G_{\rm eff}(\omega)&=&2\varsigma\Omega^2\frac{\omega \frac{2\Omega_1}{|\omega|+\Omega_1}}{\overline{\gamma}^2+(|\omega|-\Omega_1)^2},
\end{eqnarray}
where we used \mbox{$\Omega_1 n_1(0)^2=\Omega+\mathcal{O}(\alpha^2)$},
and \mbox{$\overline{\gamma}_{th}:=(2n_{th}(\Omega_1)+1)\gamma/2$}. Consequently, the dynamics of the qubit is fully determined by the knowledge of the correlation function $Q(\tau)$ and hence of the effective spectral density derived in section \ref{Jeff}.\\
We now consider the qubit dynamics for the case of the effective nonlinear bath. Therefore we determine the actual form of the correlation functions $S(\tau)$ and $R(\tau)$. From Eq. (\ref{GeffNL}) it follows:
\begin{eqnarray}
S(\tau)&=&X \tau+L\left[\exp(-\overline{\gamma}_{th}\tau)\cos\left( \Omega_1\tau\right)-1\right]\nonumber\\
&&+Z\exp(-\overline{\gamma}_{th}\tau)\sin\left(\Omega_1 \tau\right),\\
R(\tau)&=&I-\exp(-\overline{\gamma}_{th}\tau)\left[N \sin\left(\Omega_1 \tau\right)\right.\\
&&\left.+I\cos\left(\Omega_1\tau\right)\right],\nonumber
\end{eqnarray}
where
\begin{eqnarray}
I&=&\frac{2\pi\varsigma\Omega^2}{\Omega_1^2+\overline{\gamma}_{th}^2}\\
N&=&-I\left(\frac{\Omega_1}{\overline{\gamma}_{th}}-\frac{\overline{\gamma}_{th}}{\Omega_1}\right)\nonumber\\
X&=&\frac{2}{\hbar\beta}I\nonumber\\
L&=&-\frac{I}{\overline{\gamma}_{th}} \frac{1}{\cosh\left(\beta\hbar\Omega_1\right)-\cos\left(\beta\hbar\overline{\gamma}_{th}\right)}\times\nonumber\\
&&\left[\Omega_1\right.
\left.\sinh\left(\beta\hbar\Omega_1\right)-\overline{\gamma}_{th}\sin\left(\beta\hbar\overline{\gamma}_{th}\right)
\right]\nonumber
\\
Z&=&-\frac{I}{\overline{\gamma}_{th}} \frac{1}{\cosh\left(\beta\hbar\Omega_1\right)-\cos\left(\beta\hbar\overline{\gamma}_{th}\right)}\times\nonumber\\&&
\left[\overline{\gamma}_{th}\right.
\left.\sinh\left(\beta\hbar\Omega_1\right)+\Omega_1\sin\left(\beta\hbar\overline{\gamma}_{th}\right)
\right]\nonumber.\nonumber
\end{eqnarray}
Here we have neglected the contribution coming from the Matsubara term, which is verified if the temperature is high enough \cite{Weiss}, i.e. $k_B T\gg\hbar\overline{\gamma}/(2\pi)$. Moreover, we applied in the contributions of the poles lying in the vicinity of $\pm\Omega_1$ the approximation: $2\Omega_1/(2\Omega_1\pm i\overline{\gamma}_{th})\approx 1$. This corresponds effectively to neglect certain $\mathcal{O}(\overline{\gamma}_{th})$ contributions.

\subsection{Analytical solution for the nonlinear peaked spectral density}
\begin{figure}[h!]
\begin{center}
\includegraphics[width=0.45\textwidth]{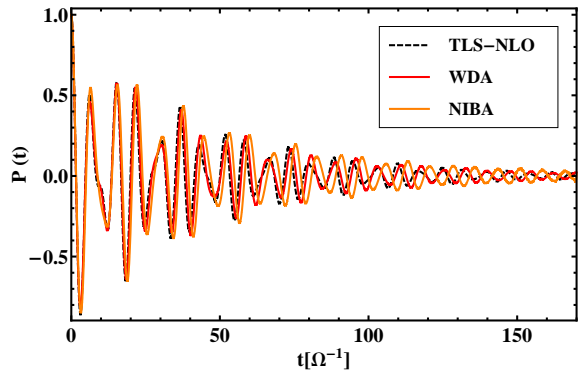}
\end{center}
\caption{Comparison of the behaviour of $P(t)$ as obtained from the numerical solution of the Bloch-Redfield equations based on the TLS-NLO approach, Ref. [\onlinecite{CarmenPRA}], the numerical solution of the NIBA equation, Eq. (\ref{GME}), and the analytical formula provided in Eq. (\ref{Panalyt}). The chosen parameters are: $\alpha=0.02(\hbar\Omega)$, $g=0.18\Omega$, $\varepsilon=0$, $\gamma/(2\pi\Omega)=0.0154$ and $\beta=10(\hbar\Omega)^{-1}$. The dynamics agree within all three approaches.
\label{Phighg}}
\end{figure}

\begin{figure}[h!]
\begin{center}
\includegraphics[width=0.45\textwidth]{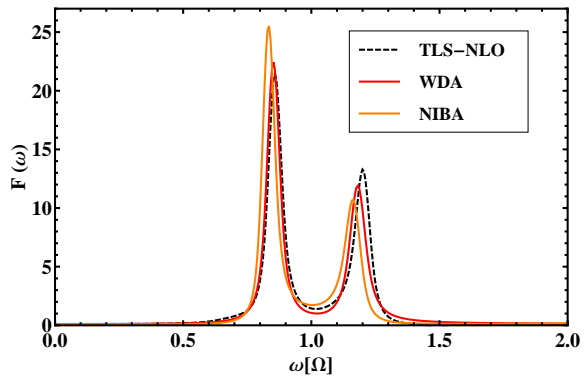}
\end{center}
\caption{Corresponding Fourier transform of $P(t)$ as shown in Fig. \ref{Phighg}.
\label{Fhighg}}
\end{figure}

\begin{figure}[h!]
\begin{center}
\includegraphics[width=0.45\textwidth]{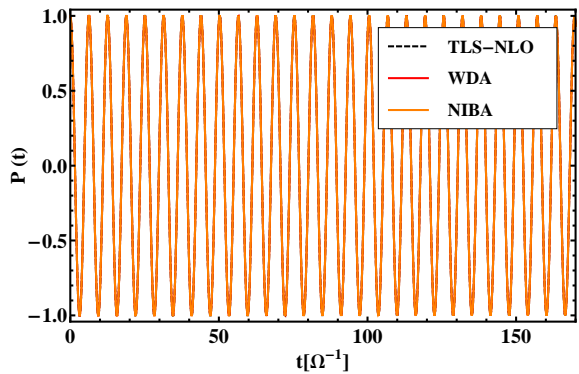}
\end{center}
\caption{As in Fig. \ref{Phighg} but for a smaller TLS-NLO coupling constant $g=0.0018\Omega$.
\label{Plowg}}
\end{figure}

\begin{figure}[h!]
\begin{center}
\includegraphics[width=0.45\textwidth]{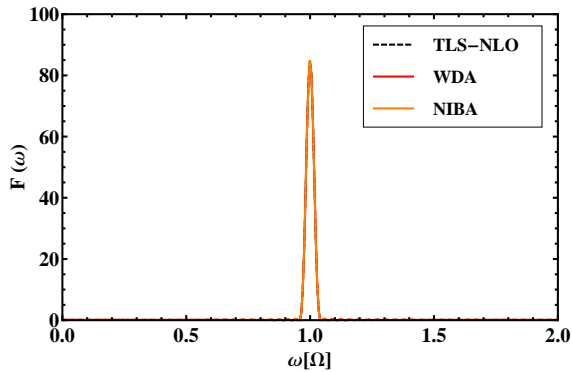}
\end{center}
\caption{Corresponding Fourier transform of $P(t)$ shown in Fig. \ref{Plowg}.
\label{Flowg}}
\end{figure}

In this section we derive an analytical formula for the population difference $P(t)$ for the symmetric case ($\varepsilon=0$), requiring weak damping strengths $\gamma$, such that a \textit{weak damping approximation} of the NIBA kernels is verified, specifically $\gamma/(2\pi\Omega)<<1$. As this calculation is analogue to the one illustrated in detail in Ref. [\onlinecite{Nesi}], we only define the relevant quantities and give the main results.\\
Due to the convolutive form of Eq. (\ref{GME}) this integro-differential equation is solved by applying Laplace transform. In Laplace space it reads:
\begin{eqnarray}
P(\lambda)&=&\frac{1-\frac{1}{\lambda}K^a(\lambda)}{\lambda+K^s(\lambda)},
\end{eqnarray}
where $P(\lambda)=\int_0^\infty dt\exp(-\lambda t)P(t)$ and analogously for $K^{a/s}(\lambda)$.\\
Consequently, the dynamics of $P(t)$ is determined if the poles of
\begin{eqnarray}\label{poleq}
\lambda+K^s(\lambda)&=&0
\end{eqnarray}
are found and the corresponding back transformation is applied.
We transform the kernels in Eq. (\ref{kernels}) in Laplace space and expand them up to first order in the damping. This procedure is called weak damping approximation (WDA) in Ref. [\onlinecite{Nesi}]. One obtains:
\begin{eqnarray}
K^{(s)}(\lambda)&=&\Delta^2\int_0^\infty d\tau \exp(-\lambda \tau)\exp(-S_0(\tau))\\
&&\left\{\cos(R_0(\tau))[1-S_1(\tau)]-\sin(R_0(\tau))R_1(\tau)\right\},\nonumber\\
K^{(a)}(\lambda)&=&0\nonumber,
\end{eqnarray}
where the indices $\{0,1\}$ denote the actual order in the damping. Specifically,
\begin{eqnarray}
S(\tau)&=&S_0(\tau)+S_1(\tau)+\mathcal{O}(\gamma^2),\\
R(\tau)&=&R_0(\tau)+R_1(\tau)+\mathcal{O}(\gamma^2),
\end{eqnarray}
where
\begin{eqnarray}
S_0(\tau)&=&Y[\cos(\Omega_1 \tau)-1],\\
S_1(\tau)&=&A\tau\cos(\Omega_1 \tau)+B\tau+C\sin(\Omega_1 \tau),\nonumber\\
R_0(\tau)&=&W\sin(\Omega_1 \tau),\nonumber\\
R_1(\tau)&=&V\left(1-\cos(\Omega_1 \tau)-\frac{\Omega_1\tau}{2}\sin(\Omega_1 \tau)\right)\nonumber.
\end{eqnarray}
The zeroth order coefficients in the damping are given by:
\begin{eqnarray}
Y&=&-W\frac{\sinh(\beta\hbar\Omega_1)}{\cosh(\beta\hbar\Omega_1)-1},\\
W&=&\frac{4 g^2n_1(0)^4}{\Omega_1\Omega(2n_{th}(\Omega_1)+1)}\nonumber,
\end{eqnarray}
and the first order coefficients by:
\begin{eqnarray}
A&=&-\overline{\gamma}_{th} Y,\nonumber\\
B&=&\frac{2}{\hbar\beta}V,\nonumber\\
C&=&-V\frac{\beta\hbar\Omega_1+\sinh(\beta\hbar\Omega_1)}{\cosh(\beta\hbar\Omega_1)-1},\nonumber\\
V&=&\frac{2g^2n_1(0)^4\gamma}{\Omega_1^2\Omega}\nonumber.
\end{eqnarray}
With this we are able to solve the pole equation for $P(t)$, Eq. (\ref{poleq}), as an expansion up to first order in the damping around the solutions $\lambda_p$ of the non-interacting pole equation, i.e., $\lambda^*=\lambda_p-\gamma \kappa_p+i\gamma\upsilon+\mathcal{O}(\gamma^2)$, as $\gamma/\Omega<<1$. Following Nesi et al. \cite{Nesi} the kernel is rewritten in the compact form:
\begin{eqnarray}
K^{(s)}(\lambda)&=&\sum_{n=0}^\infty \int_0^\infty d\tau \exp(-\lambda \tau)\left\{\Delta_{n,c}^2\cos(n\Omega_1 t)\right.\nonumber\\
&&\left.[1-S_1(\tau)]
+\Delta_{n,s}^2\sin(n\Omega_1 t)R_1(\tau)\right\},
\end{eqnarray}
where
\begin{small}
\begin{eqnarray}
\Delta_{n,c}&=&\Delta\exp(Y/2)\sqrt{(2-\delta_{n,0})(-i)^n J_n(u_0)\cosh\left(n\frac{\hbar\beta\Omega_1}{2}\right)},\nonumber\\
\Delta_{n,s}&=&\Delta\exp(Y/2)\sqrt{(2-\delta_{n,0})(-i)^n J_n(u_0)\sinh\left(n\frac{\hbar\beta\Omega_1}{2}\right)}\nonumber,\\
\end{eqnarray}
\end{small}
and
\begin{eqnarray}
u_0&=&i\sqrt{Y^2-W^2}\\
&=&i\left(\frac{4g^2n_1(0)^4}{(2n_{th}(\Omega_1)+1)\Omega_1\Omega}\right)\frac{1}{\sinh(\beta\hbar\Omega_1/2)}
.\nonumber
\end{eqnarray}
To obtain analytical expressions, we observe that in the considered parameter regime where $g/\Omega\ll1$ and $\beta\hbar\Omega_1>1$ it holds $|u_0|<1$. Following \cite{Nesi} this allows effectively a truncation to the $n=0$ and $n=1$ contributions in $K^{(s)}(\lambda)$ as the argument of the Bessel functions is small, leading to the following approximations:
\begin{eqnarray}
\Delta_{0,c}^2&=&\Delta^2\exp(Y)J_0(u_0)\approx\Delta^2\exp(Y)\\
&\approx&\Delta^2\exp\left(-\frac{4 g^2n_1(0)^6}{\Omega^2}\right)\nonumber\\
\Delta_{1,c}^2&=&\Delta^2\exp(Y)\sqrt{Y^2-W^2}\cosh(\beta\hbar\Omega_1/2),\nonumber\\
&\approx&\Delta_{0,c}^2\frac{4 g^2n_1(0)^6}{\Omega^2}.\nonumber
\end{eqnarray}
Solving the undamped pole equation yields:
\begin{eqnarray}\label{polecond}
\lefteqn{\lambda_p^2\equiv\lambda_{\pm}^2=-\frac{\Delta_{0,c}^2+\Delta_{1,c}^2+\Omega_1^2}{2}}\\&&
\pm\sqrt{\left(\frac{\Delta_{0,c}^2-\Omega_1^2}{2}\right)^2+\frac{\Delta_{1,c}^2}{2}\left(\Delta_{0,c}^2+\frac{\Delta_{1,c}}{2}^2+\Omega_1^2\right)}\nonumber\\
&:=&-\Omega_{\pm}^2.\nonumber
\end{eqnarray}
The last two equations allow to determine the oscillation frequency. Finally, within the WDA the qubit's population difference is obtained as:
\begin{small}
\begin{eqnarray}\label{Panalyt}
P(t)&=&\exp(-\gamma\kappa_- t)\frac{\lambda_-^2+\Omega_1^2}{\lambda_-^2-\lambda_+^2}\left[\cos(\Omega_- t)-\frac{\gamma\kappa_-}{\Omega_-}\sin(\Omega_- t)\right]\nonumber\\&&
+\exp(-\gamma\kappa_+ t)\frac{\lambda_+^2+\Omega_1^2}{\lambda_+^2-\lambda_-^2}\left[\cos(\Omega_+ t)-\frac{\gamma\kappa_+}{\Omega_+}\sin(\Omega_+ t)\right]\nonumber,\\
\end{eqnarray}
\end{small}
where $\kappa_{\pm}=\kappa(\lambda_\pm)$, which is derived in detail in Eq. (B.1.) of Ref. [\onlinecite{Nesi}]. Note that for a consistent treatment if $\mathcal{O}(g^2)$ is kept we implicitly require $\gamma\ll g$, as only the first order in the damping is taken into account.\\\\
We consider two possible resonance cases: First we choose the resonance condition $\Omega_1=\Delta_{0,c}$, such that the oscillation frequencies are, to lowest order in $\Delta_{1,c}$,
\begin{eqnarray}\label{freq}
\Omega_{\pm}&=&\Omega_1\mp\frac{\Delta_{1,c}}{2}\\
&\approx&\Omega+\frac{3}{\hbar}\alpha \mp g(1-\frac{3}{2\hbar}\alpha).\nonumber
\end{eqnarray}
As a consequence we obtain the so-called Bloch-Siegert shift:
\begin{eqnarray}
\Omega_--\Omega_+
&=&2 g(1-3\alpha/2\hbar),
\end{eqnarray}
which is also obtained in Ref. [\onlinecite{CarmenPRA}].
For comparison with \cite{CarmenPRA} we choose as second condition $\Delta=\Omega$, such that:
\begin{eqnarray}
\Omega_-&=&\Omega+\frac{3 \alpha}{2\hbar}-g+\frac{3 \alpha  g}{2 \hbar\Omega },\\
\Omega_+&=&\Omega+\frac{3 \alpha}{2\hbar}+g-\frac{3 \alpha g}{2 \hbar\Omega },
\end{eqnarray}
which agrees with the results of \cite{CarmenPRA} up to first order in  the nonlinearity or/and  in the coupling. Comparing with \cite{CarmenPRA} we do not observe an exact agreement for the prefactors of the mixed terms of order $\mathcal{O}(\alpha g)$.\\
We show in Figs. \ref{Phighg} and \ref{Fhighg} a comparison of the analytic WDA formula Eq. (\ref{Panalyt}), the numerical solution of the NIBA Eq. (\ref{GME}), denoted by NIBA, and the results obtained in Ref. [\onlinecite{CarmenPRA}] from a numerical solution of the Bloch-Redfield equations referred to as TLS-NLO approach. We observe that the dynamics is dominated by two frequencies and well reproduced within all three approaches. In the Fourier spectrum we observe tiny deviations of the resonance frequencies. There are two different reasons for these deviations: First the coupling strength $g$ is large enough, that higher orders in the coupling yield a finite contribution in the effective bath description. Second Eq. (\ref{polecond}) has to be expanded in both the nonlinearity and the coupling $g$, which is not possible in the numerical program. However, as we derived above when expanding the analytic formula, we find up to lowest order in the coupling $g$ and in the nonlinearity $\alpha$ the same results. We emphasize that this small discrepancy is also seen for the corresponding linear system in the work of Hausinger et al. \cite{Hannes} when comparing the NIBA results in Ref. [\onlinecite{Nesi}] with those of the Bloch-Redfield procedure.
To clarify the above statements we consider also the case that the coupling is weak, i.e., $g\ll\gamma,\Omega_1$. In the regime where the coupling is much weaker than the nonlinearity ($\hbar g\ll\alpha$), Eq. (32) given in Ref. [\onlinecite{CarmenPRA}] has to be expanded differently. Note that in this regime the results of Eq. (41) in Ref. [\onlinecite{CarmenPRA}] are not applicable. A proper expansion allows in this regime to neglect $\mathcal{O}(g^2)$ or higher if $\mathcal{O}(\alpha^2)$ is neglected. The transition frequencies, when choosing $\Omega=\Delta$, are then determined by Eq. (32) of Ref. [\onlinecite{CarmenPRA}]:
\begin{eqnarray}\label{rc1}
\Omega_{\pm}&=&\Omega+\frac{3}{2\hbar}\alpha\mp\frac{1}{2}\sqrt{9\alpha^2/\hbar^2}\\
&=&\left\{\begin{array}{c}
\Omega,\\
\Omega_1=\Omega+3\alpha/\hbar.
\end{array}
\right.\nonumber
\end{eqnarray}
Applying also an expansion of Eq. (\ref{polecond}) consistent with this parameter regime, we obtain:
\begin{eqnarray}
-\Omega_{\pm}^2&=&\frac{1}{2}\left(-\Omega^2-\Omega_1^2\pm\Omega^2\mp\Omega_1^2\right),
\end{eqnarray}
such that:
\begin{eqnarray}\label{rc2}
\Omega_{+}&=&\Omega,\\
\Omega_{-}&=&\Omega_1=\Omega+3\alpha/\hbar\nonumber.
\end{eqnarray}
The transition frequencies in Eqs. (\ref{rc1}) and (\ref{rc2}) coincide, and in Figs. \ref{Plowg} and \ref{Flowg} there is no deviation observed when comparing the three different approaches.

\subsection{Influence on the qubit dynamics due to the nonlinearity- A comparison of the NIBA for linear and nonlinear effective spectral densities}\label{Res}
\begin{figure}[h!]
\begin{center}
\includegraphics[width=0.45\textwidth]{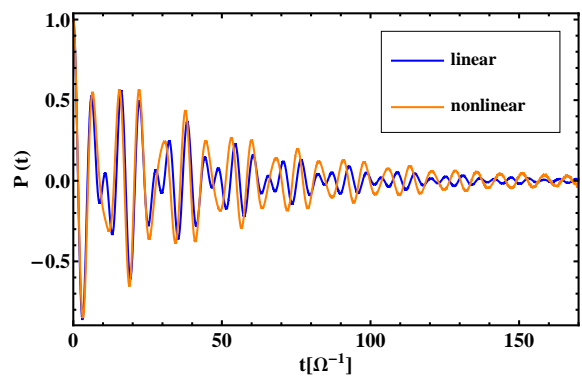}
\end{center}
\caption{$P(t)$ within the NIBA when using the linear and the nonlinear effective spectral densities, Eqs.(\ref{linearspecdens}) and (\ref{Jsimpl}) respectively. Parameters are: $\alpha=0.02(\hbar\Omega)$ or $\alpha=0$ respectively, $g=0.18\Omega$, $\varepsilon=0$, $\gamma/(2\pi\Omega)=0.0154$ and $\beta=10(\hbar\Omega)^{-1}$.
\label{CompNLLP}}
\end{figure}

\begin{figure}[h!]
\begin{center}
\includegraphics[width=0.45\textwidth]{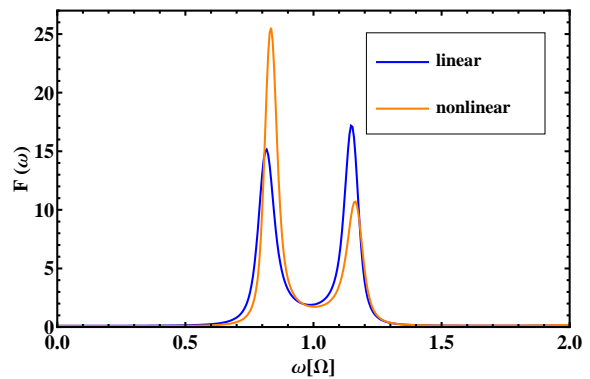}
\end{center}
\caption{Corresponding Fourier transform of $P(t)$ shown in Fig. \ref{CompNLLP}. The effect of the nonlinearity is to increase the resonance frequencies with respect to the linear case. As a consequence the relative peak heights change.
\label{CompNLLF}}
\end{figure}

In this last section we want to address the effects of the nonlinearity onto the qubit dynamics. The comparison of linear versus nonlinear case is done at level of the numerical solution of the NIBA equation and shown in Figs. \ref{CompNLLP} and \ref{CompNLLF}. As already obtained in Ref. [\onlinecite{CarmenPRA}], we observe that the transition frequencies are shifted to higher values compared to the linear case. As a consequence also the amplitudes associated to the transitions are modified. Moreover, we observe a decrease of the vacuum Rabi splitting compared to the linear case. Consequently, the effect of the nonlinearity of the read-out device can be observed in the qubit dynamics.

\section{Conclusions}\label{Conclusions}
In this work we determined the dynamics of a qubit coupled via a nonlinear oscillator (NLO) to an Ohmic bath within an effective bath description. We investigated an approximate mapping procedure based on linear response theory, which is applicable in case of weak nonlinearities and small to moderate qubit-NLO coupling. We determined the effective spectral density in terms of the qubit-oscillator coupling and the linear susceptibility of a nonlinear oscillator. The susceptibility was calculated for practical purposes from the periodically driven counterpart of the original nonlinear oscillator. The so obtained spectral density shows resonant behaviour, specifically almost a Lorentzian form for the considered parameter regime, and is peaked at a shifted frequency, namely at the one-photon resonance between ground state and first excited state of the nonlinear oscillator. Moreover, this effective spectral density acquires a temperature dependence and behaves Ohmic at low frequencies. Based on the effective spectral density the qubit dynamics are investigated within the NIBA approximation. In addition an analytical formula for the qubit dynamics is provided, which describes very well the dynamics at low damping. These results were compared to the numerical solution in Ref. [\onlinecite{CarmenPRA}], where the Bloch-Redfield equations for the density matrix of the coupled qubit nonlinear oscillator system (TLS-NLO) are solved. We find an overall agreement of the two approaches and show that deviations are of order $\mathcal{O}(\alpha g)$, where $\alpha$ is the nonlinearity and $g$ the coupling strength. Exemplarily this effect was analyzed for two possible coupling strengths $g$. We emphasize that parameters like temperature and damping and especially the strength of the coupling $g$ and nonlinearity $\alpha$ determine the appropriate form of the expansions in the different regime of parameters. Due to the tunability of the parameters various qubit dynamics are possible. In agreement with Ref. [\onlinecite{CarmenPRA}] we observed the following effects due to the nonlinearity: First the transition frequencies of the two dominating peaks, where we are in the regime $g\gg\alpha/\hbar,$ are shifted to larger values compared to the linear case. As a consequence also the amplitudes of the coherent oscillations of the population difference $P(t)$ are modified. Moreover, the Bloch-Siegert shift is decreased due to the nonlinearity.\\
We conclude that, as in case of the corresponding linear system \cite{Nesi,Hannes}, the effective bath description provides an alternative approach to investigate the complex dynamics of the qubit dissipative NLO system.
\section{Acknowledgment}
We acknowledge support by the DFG under the funding program SFB 631.


\end{document}